\documentclass[amsmath,amssymb,
aps,
citeautoscript,superscriptaddress,twocolumn,showpacs,10pt]{revtex4-1}

\usepackage{bm}
\usepackage{graphicx}
\usepackage{dcolumn}
\usepackage[colorlinks,linkcolor=red,anchorcolor=green,
citecolor=blue,breaklinks]{hyperref}

\usepackage{mathrsfs}
\usepackage{color}
\usepackage{braket}
\usepackage[normalem]{ulem}

\graphicspath{{figures/}}
\newcommand{\HF}{\text{HF}}

\newcommand{\blambda}{\boldsymbol{\lambda}}

\newcommand{\ha}{\hat{a}}

\newcommand{\hH}{\hat{H}}

\begin{document}

\preprint{}

\title{First-principles molecular quantum electrodynamics theory at all coupling strengths}

\author{Xinyang Li}
\affiliation{
Theoretical Division, Los Alamos National Laboratory, Los Alamos, New Mexico 87545, United States}

\author{Yu Zhang}
\email[]{zhy@lanl.gov}
\affiliation{
Theoretical Division, Los Alamos National Laboratory, Los Alamos, New Mexico 87545, United States}

\date{\today}

\begin{abstract}
The ever-growing intersection of quantum electrodynamics (QED) and molecular processes has shown remarkable and unanticipated advancements in altering molecular properties and reactivity by exploiting light-matter couplings. In recent years, multiple ab initio methods have been developed to compute the eigenstates of molecular systems strongly coupled to cavities, ranging from the mean-field to quantum many-body methods. The quantum many-body methods, such as coupled-cluster theories, usually rely on the quality of mean-field reference wavefunctions. Hence, developing efficient and physically reliable mean-filed approaches for molecular quantum electrodynamics problems is crucial. The current widely used methods, such as QED Hartree-Fock and the self-consistent counterpart, are limited to specific coupling regimes. In this work, we developed a variational transformation-based molecular quantum electrodynamics mean-field method, namely VT-QEDHF, for light-matter interaction at arbitrary coupling strength. The numerical benchmark demonstrates that the VT-QEDHF method naturally connects both QEDHF and self-consistent QEDHF methods at the two limits, showcasing the advantage of VT-QEHDF across all coupling strengths. 
\end{abstract}

\maketitle

\section{Introduction.}
The increasing overlap between quantum electrodynamics (QED) and molecular activities has led to breakthroughs in tailoring molecular properties and activities through light-matter interactions~\cite{Ebbesen2016ACR, Mandal:2023vh, Weight2023pccp}. When strongly coupled, both photons and electrons (or other elementary excitations) within materials become essential and intermingle equally quantized. In such an environment, the concept of independent ``free'' particles ceases to exist. Instead, the elementary excitations in the strong light-matter interaction regime are polaritons, which represent a superposition between quantized light and material~\cite{Revera2020NatRevPhys} and display characteristics of both light and matter. Research suggests that material properties can be modulated via these polaritons, engendering a diversity of photophysical and photochemical phenomena; that is, polariton chemistry~\cite{Ebbesen2016ACR}. Given that the energies of photons and the strength of light-matter interactions can be fine-tuned through cavity manipulations, the robust coupling between light and matter unveils a novel paradigm for modifying material characteristics, with a spectrum of possible applications including lasing~\cite{Kena-Cohen:2010ue, Kang:ws}, long-distance energy transmission~\cite{Zhong:2017vq, Georgiou2021angwew, Want2021natcomm, Coles:2014wm, timmer_plasmon_2023}, Bose-Einstein condensates~\cite{Dusel:2020wt, Zasedatelev:2019um, Kavokin:2022tn}, and various chemical processes~\cite{Pavosevic2023, Pavosevic2022b, Schaefer2022NC, Schaefer2022JPCL, Cave1997JCP, MartinezMartinez2017AP, Yang2021JPCL, Climent2020PCCP, Wang2022JPCL, Imperatore2021JCP, Galego2016NC, CamposGonzalezAngulo2019NC, Philbin2022JPCC, Efrima1974CPL, Phuc2020SP, Davidsson2020JCP, Galego2017PRL, Mauro2021PRB, Vurgaftman2020JPCL, Hiura2021C, Hiura2019-rate, Weight2023}.

In the thriving field of polariton chemistry (or molecular quantum electrodynamics at large), investigating the influence of arbitrary light-matter coupling strengths on molecular properties and behaviors necessitates a robust and universally applicable theoretical approach~\cite{Fregoni2022acsphotonics}. However, the absence of a reliable theoretical framework that seamlessly traverses all coupling regimes hinders the full potential of QED-assisted modulation of molecular properties. Despite significant progress in understanding the effects of confined fields on many molecular characteristics, a comprehensive and first-principles framework for exploring these phenomena across all coupling regimes is still lacking. To date, variational theories~\cite{Rivera:2019tr}, QED Hartree-Fock (QEDHF)~\cite{haugland_coupled_2020, Riso:2022uw}, semi-empirical method~\cite{zhang2019jcp}, QED Density Functional Theory (QED-DFT)~\cite{Flick2017PNAS, Ruggenthaler2014PRA, Schaefer2021}, QED coupled cluster (QED-CC)~\cite{haugland_coupled_2020, jcp0033132, Liebenthal2022, mordovina_polaritonic_2020, Weight2023pccp}, QED Time-Dependent Density Functional Theory (QED-TDDFT)~\cite{yang_quantum-electrodynamical_2021}, and Diffusion Quantum Monte Carlo~\cite{weight_diffusion_2023} methods have been proposed to study the light-matter interactions. In particular, post-Hartree-Fock methods depend on an optimal mean-field theory (as the reference state) to achieve better accuracy. Although they are effective in addressing several aspects of molecular interactions within quantum fields, the existing QEDHF methods~\cite{haugland_coupled_2020} and their self-consistent counterparts~\cite{Riso:2022uw} are primarily limited to specific coupling strengths.

To address this research gap, we introduce a variational transformation~\cite{Zhang:2015ua} based first-principles QED method, referred to as the VT-QEDHF. This universal approach is designed to function effectively across arbitrary coupling strengths, thereby providing an invaluable tool for exploring and understanding light-matter interactions in a more comprehensive and efficient manner. The VT-QEDHF method transcends the limitations of traditional perturbative and strong coupling approaches, offering a more universal perspective on molecular processes in QED environments. Within the VT-QEDHF framework, the photonic field contribution is accounted for in a nonperturbative manner, ensuring the attainment of the exact wave function in the limit of infinite coupling, thereby providing a consistent and reliable molecular orbital description across various coupling regimes. This first-principles approach not only captures the electron-photon correlation (at the mean-field level) effectively but also elucidates the cavity effects on the electronic ground state while maintaining a manageable computational cost. By bridging the theoretical gap across coupling strengths, the VT-QEDHF method is anticipated to open new avenues for the study and manipulation of molecular properties and behaviors within QED environments, offering enriched insights and enhancing the predictability and control over light-matter interactions.

\section{Theory}
The total light-matter Hamiltonian of molecular quantum electrodynamics can be described as the widely used nonrelativistic Pauli-Fierz Hamiltonian in the dipole approximation~\cite{CohenTannoudji1997, Mandal:2023vh, Weight2023pccp},
\begin{align}\label{EQ:H_PF}
    \hat{H}_\mathrm{PF}
     =& \hat{H}_\mathrm{e} + \sum_{\alpha} \Big[ \omega_\alpha (\hat{a}^\dagger_\alpha\hat{a}_\alpha+\frac{1}{2})
    \nonumber\\
    &+\sqrt{\frac{\omega_\alpha}{2}} \boldsymbol{\lambda}_\alpha \cdot \hat{\boldsymbol{D}} (\hat{a}^\dagger_\alpha + \hat{a}_\alpha) + \frac{1}{2} (\boldsymbol{\lambda}_\alpha \cdot \hat{\boldsymbol{D}})^2 \Big].
\end{align}
This Hamiltonian is often referred to as the Pauli-Fierz (PF) Hamiltonian.
Where $\hat{H}_\mathrm{e} = \hat{T}_\mathrm{e} + \hat{V}$ is the bare molecular Hamiltonian (excluding the nuclear kinetic operator) which includes all Coulomb interactions $\hat{V}$ between electrons and nuclei as well as the electronic kinetic energy operators $\hat{T}_\mathrm{e}$, which is given by the expression,
\begin{equation}
    \hat{H}_e = \sum_{\mu\nu}h_{\mu\nu} \hat{c}^\dagger_\mu \hat{c}_\nu +
    \frac{1}{2}\sum_{\mu\nu\lambda\sigma}I_{\mu\nu\lambda\sigma} \hat{c}^\dagger_\mu \hat{c}^\dagger_\lambda \hat{c}_\sigma \hat{c}_\nu. 
\end{equation}
Where $h$ and $I$ are one-electron and two-electron integrals.  $\hat{\boldsymbol{D}}$ in Eq.~\ref{EQ:H_PF} is the molecular dipole operator,
\begin{equation}
\hat{\boldsymbol{D}}=\sum_i^{N_n} z_i\hat{\boldsymbol{R}}_i -\sum_i^{N_e} e\hat{\boldsymbol{r}}_i
\equiv\hat{\boldsymbol{D}}_n + \hat{\boldsymbol{D}}_e,
\end{equation}
including electronic $\hat{\boldsymbol{D}}_e$ and nuclear $\hat{\boldsymbol{D}}_n$ components. $\boldsymbol{\lambda}_\alpha =\sqrt{\frac{1}{\epsilon_0 V}}\boldsymbol{e}_\alpha \equiv \lambda_\alpha \boldsymbol{e}_\alpha$ characterizes the coupling between the molecule and cavity quantized field. $\omega_\alpha$ and $\boldsymbol{e}_\alpha$ represent the frequency and polarization of the electric field of cavity photon mode $\alpha$. The last term describes the dipole self-energy (DSE), which is essential to ensure the Hamiltonian is bounded from below and displays the correct scaling with the system size~\cite{Rokaj2018JPBAMOP}.

The eigenstate of the molecular QED Hamiltonian can be readily obtained by solving the time-independent Schr\"odinger equation
\begin{equation}
    \hat{H}_{\mathrm{PF}}\ket{\Psi} = E\ket{\Psi},
\end{equation}
where $\ket{\Psi}$ is the correlated electron-photon wavefunction, though the exact solution to the above quantum many-body equation is nontrivial.

The mean-field approach is usually the first and fastest method to approximate the quantum many-body problems.
At the mean-field level, the QED Hamiltonian can be approximated by $\ket{\Psi}\approx \ket{\mathrm{HF}}\otimes\ket{0}$ where $\ket{0}$ denotes the photon vacuum state. Consequently, the total energy can be easily introduced via,
\begin{equation}
    E_{\mathrm{tot}} = E_{HF} + \frac{1}{2} \sum_\alpha \langle (\boldsymbol{\lambda}_\alpha \cdot \hat{\boldsymbol{D}})^2\rangle.
\end{equation}
Where the $E_{HF}$ denotes the electronic HF energy. The DSE contribution to the total energy (second term on the right-hand side of the above equation) can be evaluated via the DSE-mediated one-electron and two-electron integrals (see more details in Supplementary Materials (SM)). Thus, the corresponding Fock matrix (for computing density matrix and molecular orbital properties) can be readily derived by taking the partial derivative of the total energy with respect to the density matrix~\cite{haugland_coupled_2020, foley:2023CPR}. The resulting QEDHF method (in the Fock state representation) provides an economical way to compute the polariton ground state and can serve as the reference for other post-HF methods. The key drawback of the QEDHF method in the Fock representation is the slow convergence with the Fock state in the strong coupling limit, which can lead to incorrect behavior, such as incorrect origin-dependency and frequency dependency~\cite{foley:2023CPR}, making the QEDHF method in Fock state representation more suitable for weak coupling systems (as the Fock state is the eigenstate of the interaction Hamiltonian in the $\lambda\rightarrow 0$ limit). Such drawbacks can be mitigated with the coherent state (CS) representation~\cite{ajp4876963},
\begin{equation} 
\ket{z_\alpha}\equiv e^{z_\alpha\hat{a}_\alpha^\dagger - z_\alpha^*\hat{a}_\alpha}\ket{0}\equiv \hat{U}(z_\alpha)\ket{0},
\end{equation}
where $z_\alpha =-\frac{\langle \boldsymbol{\lambda}_\alpha \cdot \hat{\boldsymbol{D}}\rangle_{\mathrm{HF}}}{\sqrt{2\omega_\alpha}}$. It's clear from the above equation that CS is a linear combination of complete Fock states where the coefficients are determined by the displacement due to the light-matter coupling strength. The resulting QEDHF in CS representation~\cite{haugland_coupled_2020} thus mitigates the origin-variance problem. However, the molecular orbitals and Fock matrix remain origin-dependent charged systems in~\cite{foley:2023CPR}.

Only recently, a fully origin-invariant formulation was developed within a self-consistent strong coupling QEDHF formalism (namely SC-QEDHF). The SC-QEDHF framework is stimulated by the fact that, in the infinite coupling limit (i.e., $\hat{H}_e \ll \hat{H}_\mathrm{p} + \hat{H}_\mathrm{ep}$ or $\lambda_\alpha\rightarrow \infty$), the Hamiltonian is dominated by the photon and electron-photon interaction terms, and the corresponding wavefunction can be well approximated by a Gaussian state,
\begin{equation}\label{eq:gaussian}
    \ket{\Psi^{\infty}}=e^{-\sum_\alpha\frac{\lambda_\alpha}{\sqrt{2\omega_\alpha}}\boldsymbol{e}_\alpha\cdot\hat{\boldsymbol{D}}(\hat{a}_\alpha - \hat{a}^\dag_\alpha)}\ket{\mathrm{HF},0}
    \equiv \hat{U}_\lambda \ket{\mathrm{HF},0}.
\end{equation}
This is widely recognized as the polaron transformation within the context of electron-phonon interaction scenarios~\cite{Shneyder2020prb, Zhang:2015interference, mah00}. This approach has recently been adapted for use in polariton chemistry~\cite{Mandal2020JPCL, Riso:2022uw, Ashida2021PRL}. Consequently, we can employ the $\hat{U}_\lambda$ operator to transpose the Hamiltonian into a new framework, wherein the resultant transformed Hamiltonian effectively eliminates the explicit electron-photon coupling terms. In particular, after undergoing the transformation, the electronic and photonic operators become
\begin{align}
    \hat U^\dag_\lambda \hat c_\nu \hat U_\lambda = &\sum_\mu \hat c_\mu X_{\mu\nu}, \\
    \hat U^\dag_\lambda \hat a_\alpha \hat U_\lambda = & \hat a_\alpha -  \frac{\lambda_\alpha}{\sqrt{2\omega_\alpha}} \boldsymbol{e}_\alpha\cdot\hat{\boldsymbol{D}},
\end{align}
where $X_{\mu\nu} =\exp\left[-\sum_{\alpha}\frac{\lambda_\alpha}{\sqrt{2\omega_\alpha}} \boldsymbol{e}_\alpha\cdot\hat{\boldsymbol{D}}(\hat a^\dag_\alpha - \hat a_\alpha) \right]|_{\mu\nu}$.

Consequently, under the polariton transformation, the resulting Hamiltonian becomes (denoted as $\hat{H}^p$)
\begin{equation}\label{eq:polaritonh}
    \hat{H}^p = \hat{U}^\dag_\lambda  \hat{H}_{\mathrm{PF}} \hat{U}_\lambda = \hat{U}^\dag_\lambda \hat{H}_{e} \hat{U}_\lambda + \sum_\alpha \omega_\alpha\hat{a}^\dag_\alpha\hat{a}_\alpha.
\end{equation}
The transformed electronic Hamiltonian $\tilde{H}_e \equiv  \hat{U}^\dag_\lambda \hat{H}_{e} \hat{U}_\lambda$ is formally the same as the original one with the electronic operators dressed by the $X$ operator. Since the dipole coupling operator $\boldsymbol{e}_\alpha\cdot\hat{\boldsymbol{D}}$ in the $X$ operator is not diagonal, it's more convenient to transform the operator into the dipole basis (defined as the eigenstate of $\boldsymbol{e}_\alpha\cdot\hat{\boldsymbol{D}}$ operator, denoted by the symbols $p, q, r, s$ and the corresponding eigenvalues are denoted as $\eta_p$). Then, the corresponding QEDHF energies and Fock matrix can be derived. More details can be found in Ref.~\cite{Riso:2022uw}.

To bridge the treatment in weak and strong coupling limits, here we present a variational transformation-based QEDHF method for the arbitrary coupling regime. The central idea is that, instead of using $\hat{U}_\lambda$, we adopt variational parameters $f_\alpha$ to control the variational transformation  $\hat{U}_f$~\cite{Silbey1984jcp} (also called Lang-Firsov transformation~\cite{lang1963kinetic})
\begin{equation}
  \hat{U}_f = e^{-\sum_\alpha \frac{f_\alpha}{\sqrt{2\omega_\alpha}} \boldsymbol{e}_\alpha\cdot\hat{\boldsymbol{D}}(\hat{a}_\alpha-\hat{a}^\dag_\alpha)}.
\end{equation}
which helps the seek for an optimal mean-field approximation to the cavity QED Hamiltonian. Such idea has been previously used in strong electron/exciton-phonon interactions, including exciton transport~\cite{Silbey1984jcp}, polaron formation~\cite{Bari:2002prb, Alex:1994prb, Luo_polaron:2022prb}, and dissipative quantum transport~\cite{Zhang:2015ua, Hsieh2019jpcc, Wang:2020prb}.

With the variational transformation (VT), the resulting Hamiltonians become,
\begin{align}\label{eq:varh}
  \hat{H}(\{f_\alpha\}) = & \tilde{H}_e(\{f_\alpha\}) + \sum_{\alpha} \omega_{\alpha}\hat{a}^\dag_\alpha \hat{a}_\alpha
  \nonumber \\
  & + \sum_{\alpha}\sqrt{\frac{\omega_\alpha}{2}}(\Delta\lambda_\alpha) \boldsymbol{e}_\alpha\cdot\hat{\boldsymbol{D}}(\hat{a}^\dag_\alpha+\hat{a}_\alpha )
  \nonumber \\
  & + \frac{(\Delta\lambda_\alpha)^2}{2}(\boldsymbol{e}_\alpha\cdot\hat{\boldsymbol{D}})^2.
\end{align}
where $\Delta\lambda_\alpha=\lambda_\alpha - f_\alpha$ and the parameters $f_\alpha$ are to be variationally minimized. $\tilde{H}_e(\{f_\alpha\})$ is the VT dressed electronic Hamiltonian, where the original electronic operator becomes $\hat{U}^\dag_f \hat{c}_\nu \hat{U}_f = \sum_\nu \hat{c}_\nu X^f_{\mu\nu}$ and $X^f_{\mu\nu} =\exp\left[-\sum_{\alpha}\frac{f_\alpha}{\sqrt{2\omega_\alpha}} \boldsymbol{e}_\alpha\cdot\hat{\boldsymbol{D}}(\hat{a}^\dag_\alpha - \hat{a}_\alpha) \right]|_{\mu\nu}$.
The detailed derivation can be found in Supplementary Materials (SM).  Compared to the fully transformed polariton Hamiltonian in Eq.~\ref{eq:polaritonh}, the variationally transformed Hamiltonian in Eq.~\ref{eq:varh} includes a partially dressed electronic Hamiltonian $\tilde{H}_e(\{f_\alpha\})$ and residues in the bilinear coupling and DSE terms (controlled by $f_\alpha$). The last two terms in Eq.~\ref{eq:varh} are referred to as the residual bilinear coupling and DSE terms, respectively. It's obvious that when $f_\alpha/\lambda_\alpha=0$ (or 1), Eq.~\ref{eq:varh} reduces to the original PF Hamiltonian $\hat{H}_{PF}$ or fully transformed polariton Hamiltonian $\hat{H}^p$. It should be noted that the transformed Hamiltonians in Equations~\ref{eq:varh} and ~\ref{eq:polaritonh} are both exact, as no approximation was made in the transformation. The exact diagonalization of the two Hamiltonians should give the same eigenstates.

Applying the mean-field approximation to the wavefunction allows us to define the VT-QEDHF wave function as
\begin{equation}
    \ket{\Psi} = e^{-\frac{f_\alpha}{\sqrt{2\omega_\alpha}} \boldsymbol{e}_\alpha \cdot \hat{\boldsymbol{D}} (\hat{a}_\alpha - \hat{a}^\dag_\alpha)} \ket{\mathrm{HF}, 0}
    \equiv \hat{U}_f \ket{\mathrm{HF}, 0}.
\end{equation}
In the dipole basis~\cite{Riso2022}, this becomes
\begin{equation}
    \ket{\Psi} = e^{-\frac{f_\alpha}{\sqrt{2\omega_\alpha}} \sum_p \eta_p \hat{c}^\dag_p \hat{c}_p (a_\alpha - a^\dag_\alpha)} \ket{\mathrm{HF}, 0},
\end{equation}
and the transformed electronic operators in the dipole basis are given by
\begin{equation}
    \hat{U}_f^\dag \hat{c}_p \hat{U}_f = \sum_\nu \hat{c}_p X^f_{p},
\end{equation}
where \( X^f_{p} = \exp \left[-\sum_{\alpha} \frac{f_\alpha}{\sqrt{2\omega_\alpha}} (\boldsymbol{e}_\alpha \cdot \hat{\boldsymbol{D}})_{pp} (\hat{a}^\dag_\alpha - \hat{a}_\alpha) \right] \).

Consequently, the VT-QEDHF energy in the dipole basis becomes
\begin{widetext}
\begin{align}\label{eq:vtqedhfe}
    E = & \sum_{pq} \tilde{h}_{pq} \rho_{pq} G_{pq}
    + \frac{1}{2} \sum_{pqrs} \tilde{I}_{pqrs} \left( \rho_{pq} \rho_{rs} - \frac{1}{2} \rho_{ps} \rho_{rq} \right) G_{pqrs} + \frac{f^2_\alpha}{2} \sum_p \rho_{pp} \left[ (\boldsymbol{e}_\alpha \cdot \boldsymbol{D})_{pp} - \eta_p \right]^2 \nonumber \\
    & + \frac{f^2_\alpha}{2} \sum_{pq} \left( \rho_{pp} \rho_{qq} - \frac{1}{2} \rho_{pq} \rho_{qp} \right) \left[ (\boldsymbol{e}_\alpha \cdot \boldsymbol{D})_{pp} - \eta_p \right] \left[ (\boldsymbol{e}_\alpha \cdot \boldsymbol{D})_{qq} - \eta_q \right] + \frac{(\Delta\lambda_\alpha)^2}{2} \bra{\mathrm{HF}}\bra{0} (\boldsymbol{e}_\alpha \cdot \hat{\boldsymbol{D}})^2 \ket{\mathrm{HF}}\ket{0}.
\end{align}
\end{widetext}
Here, $\tilde{h}$ and $\tilde{I}$ represent one-electron and two-electron integrals in the dipole basis, respectively, $\rho_{pq}$ is the density matrix, and $G$ are the Franck-Condon factors derived by integrating out the photonic degrees of freedom from the VT-dressed one-/two-electron integrals (i.e., $\bra{0} (X^f)^\dag_p X^f_q \ket{0}$~\cite{Zhang:2015ua, Zhang:2015interference}). The first two terms in Eq.~\eqref{eq:vtqedhfe} are formally the same as the HF energy of the pure electronic system, but with one-/two-electron integrals replaced by the VT-dressed ones. The third and fourth terms account for relaxation in the dipole basis set~\cite{Riso:2022uw}. Finally, the last term in Eq.~\eqref{eq:vtqedhfe} represents the residual DSE.

The explicit form of $G$ can be found in the Supplementary Material (SM). The corresponding Fock matrix can be derived from the energy derivatives with respect to the density matrix. Moreover, the optimal $\{f_\alpha\}$ can also be optimized during the SCF procedure via the energy derivatives with respect to $f_\alpha$ (i.e., $\frac{\partial E}{\partial f_\alpha}$). The detailed formulas for the Fock matrix and $\frac{\partial E}{\partial f_\alpha}$, which are used for updating the density matrix and variational parameters, can be found in the SM. Additionally, VT-QEDHF can be augmented with the CS basis set, defined by the residue bilinear coupling as $z^f_{\alpha} \equiv -\frac{f_\alpha \langle \boldsymbol{e}_\alpha \cdot \hat{\boldsymbol{D}} \rangle}{\sqrt{2\omega_\alpha}} = -\frac{f_\alpha}{\lambda_\alpha} z_\alpha$, leading to the effective ansatz
\begin{equation}
    \Psi = e^{-\sum_{\alpha p} \frac{f_\alpha}{\sqrt{2\omega_\alpha}} \eta_p \hat{c}^\dag_p \hat{c}_p (a_\alpha - a^\dag_\alpha)} \hat{U}(z^f_{\alpha}) \ket{\mathrm{HF}}\ket{0}.
\end{equation}
This resulting formalism is denoted as the VT-QEDHF-CS method.

\begin{figure}[!htb]
    \centering
    \includegraphics[width=0.485\textwidth]{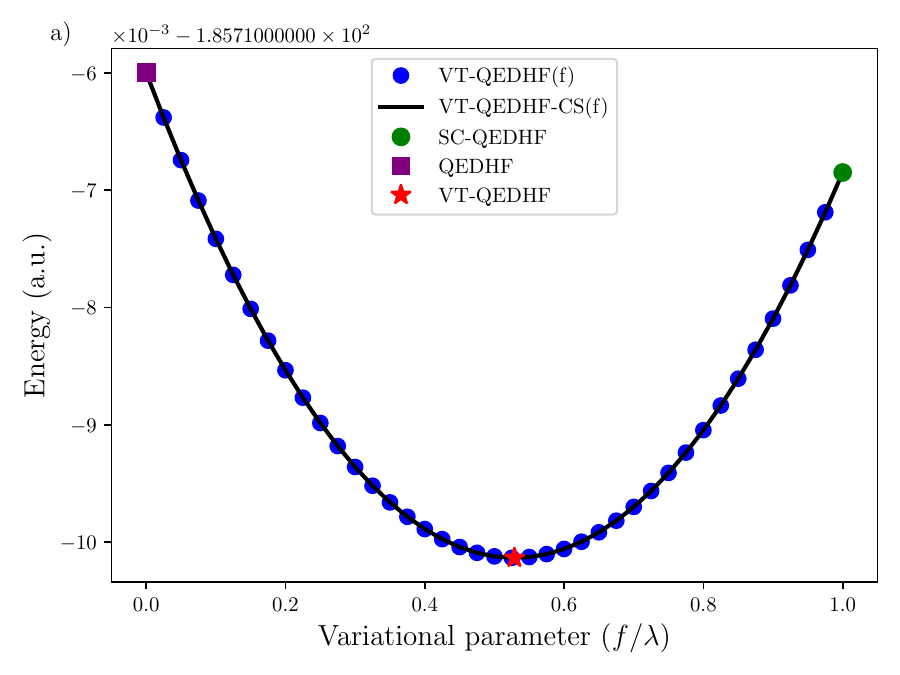}
    \includegraphics[width=0.485\textwidth]{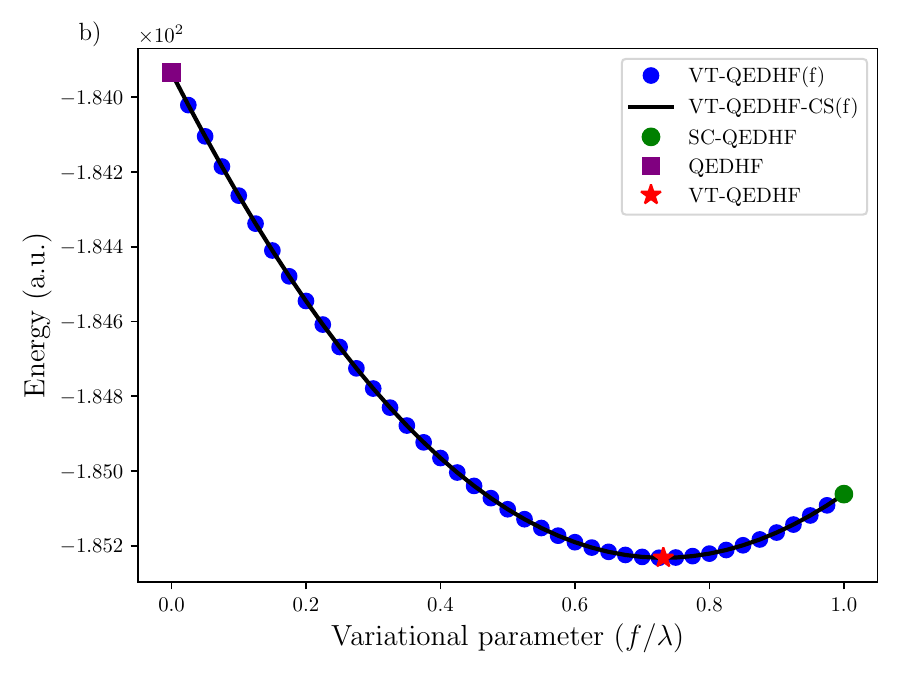}
    \caption{VT-QEDHF energies as a function of the transformation parameter $f$, showing a natural connection to QEDHF and SC-QEDHF methods at the two limits, i.e., with ($f=1$, red dot) and without ($f=0$, purple square) polariton transformation. The photon frequency $\omega$ is set to 0.5~au. The coupling strengths $\lambda$ are a) 0.05 and b) 0.5, respectively.}
    \label{fig:vscqed_vs_f}
\end{figure}

\section{Numerical Examples} 
We demonstrate the validity and advantages of the VT-QEDHF method across various coupling strengths using a sample molecule (C$_2$N$_2$H$_6$ isomer, with the STO-3G basis set employed). Configurations of the isomer along the trans-cis pathway are detailed in the Supplementary Material (SM). Figure~\ref{fig:vscqed_vs_f} plots the ground state energy of the C$_2$N$_2$H$_6$ molecule using different methods. The VT-QEDHF method with a predefined variational parameter $f$ (i.e., without optimizing $f$) is referred to as the VT-QEDHF(f) method. This method shows a natural progression to the QEDHF and SC-QEDHF methods at the limits of $f=0$ and $f=\lambda$, respectively. The red star in Figure~\ref{fig:vscqed_vs_f} indicates the optimized VT-QEDHF energy, which is the lowest among the VT-QEDHF(f) energies as shown. The optimized $f$ values for the weaker (Figure~\ref{fig:vscqed_vs_f}a) and stronger (Figure~\ref{fig:vscqed_vs_f}b) coupling cases are 0.53 and 0.73, respectively. These values suggest that stronger couplings necessitate greater transformation in the Hamiltonian, with the corresponding results more closely aligned with the SC-QEDHF method.

Furthermore, the additional optimization of $f$ does not notably amplify the SCF optimization workload. For the calculations in Fig~\ref{fig:vscqed_vs_f}, the SC-QEDHF method reaches convergence after 26 iterations, while the VT-QEDHF method meets the same criteria after 36 iterations, indicating a marginal increase in computational duration. Although the VT-QEDHF method incorporates both VT-dressed and DSE-mediated one-/two-electron integrals, the computation of the VT-dressed one-electron and two-electron integrals predominantly contributes to the bottleneck. This computation must be undertaken in every iteration, which is the same in the SC-QEDHF method. Consequently, the computational expenses of the VT-QEDHF and scQEDHF methods are nearly equivalent.

\begin{figure}
    \centering
    \includegraphics[width=0.5\textwidth]{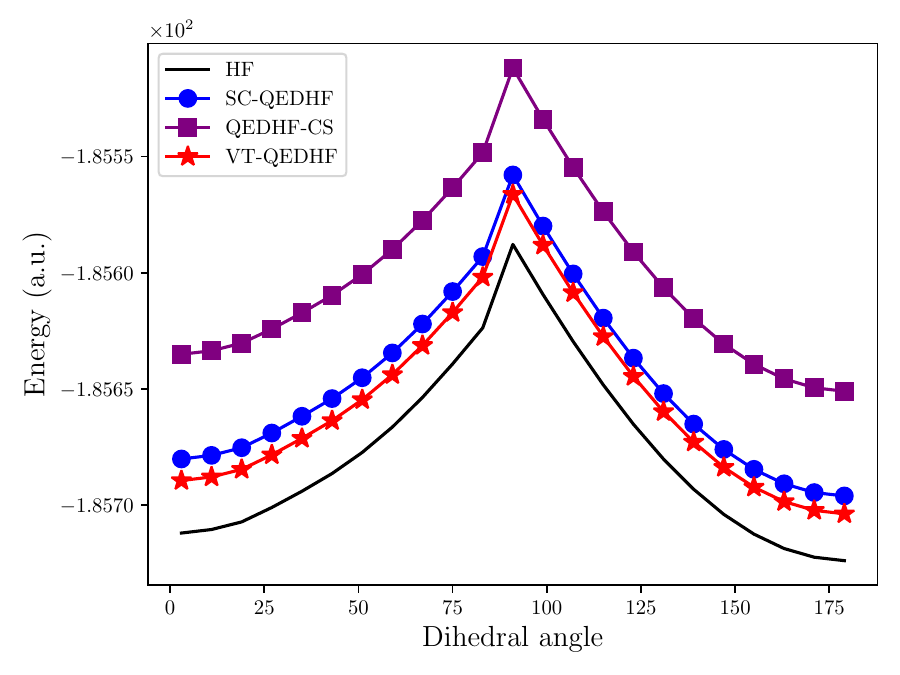}
    \caption{Ground state potential energy surfaces of C$_2$N$_2$H$_6$ isomer calculated from different methods. The photon frequency and coupling strength ($\lambda$) are 0.1 and 0.5~au, respectively.}
    \label{fig:pes}
\end{figure}

Subsequently, we determined the polariton ground state energies along the trans-cis reaction pathway using the HF and QEDHF methods. These results are depicted in Fig.~\ref{fig:pes}, with the photon frequency and coupling parameter ($\lambda$) set at 0.1 and 0.5~au, respectively. Compared with the QEDHF and SC-QEDHF methods, VT-QEDHF captures a larger amount of electron-photon correlation. This leads to reduced ground state energies throughout the reaction pathway, underscoring its reliable performance along the reaction coordinate.

\begin{figure}[htb]
    \centering
    \includegraphics[width=0.5\textwidth]{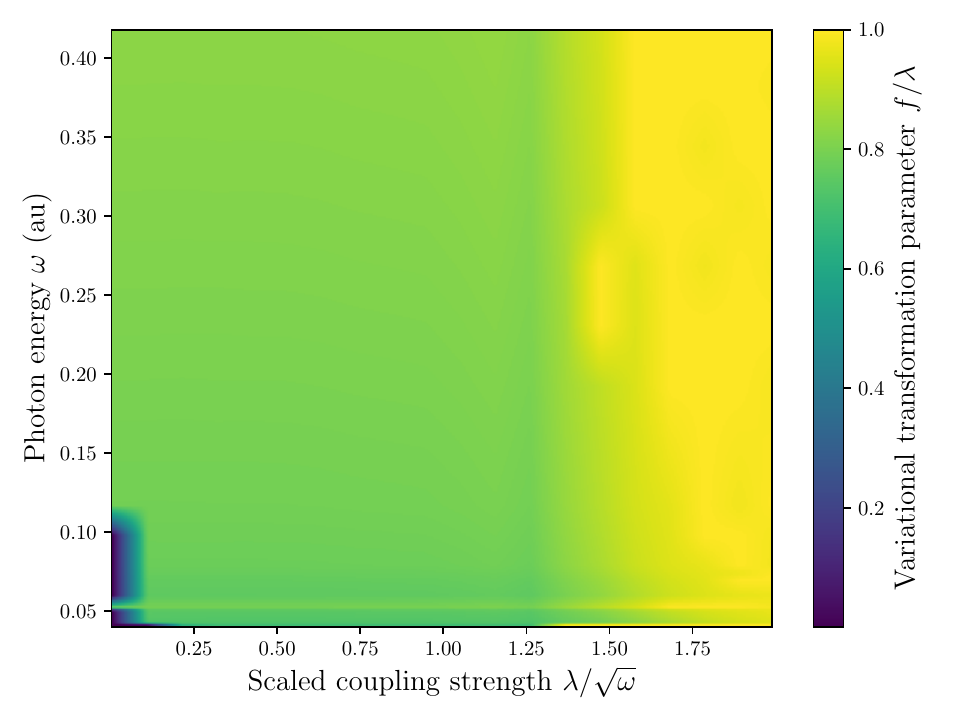}
    \caption{The relationship between optimized variational transformation parameters ($f/\lambda$) and photon frequencies, with respect to scaled coupling strengths ($\lambda/\sqrt{\omega}$). The parameter $f/\lambda$ trends towards 0 and 1 in the regimes of weak and strong coupling, respectively.}
    \label{fig:2d}
\end{figure}

We investigated the optimal variational transformation $f$ across varying photon frequencies and electron-photon coupling strengths. The LiH molecule is used here to scan a wide parameter space efficiently. These results are illustrated in Fig.~\ref{fig:2d}. As anticipated, varying electron-photon coupling strengths dictate distinct optimal values for $f$ in the variational transformation. Moreover, $f$ displays a consistent increase with the electron-photon coupling strength $\lambda$. As $\lambda$ tends toward small values, the ratio $f/\lambda$ gravitates towards zero or a finite value contingent on photon energies.
Nevertheless, in the weak coupling scenario where $\lambda \rightarrow 0$, the $f/\lambda$ ratio remains low, aligning with a minimal (or no) polariton transformation limit. Conversely, the $f/\lambda$ ratio is near unity in the strong coupling domain, reflecting a comprehensive polariton transformation. Within the intermediate range, the variational transformation culminates with a finite value for $f/\lambda$. This highlights the imperative nature of the variational transformation across a broad parameter regime to obtain optimal mean-field ground states. 

\section{Summary}
In summary, this study introduces the variational transformation-based electronic structure theory (VT-QEDHF) for molecular QED applications encompassing all ranges of coupling strengths. This methodology adeptly captures the optimal mean-field part of both electron-photon and photon-mediated electron-electron correlations. Furthermore, this framework is universally applicable to any fermion-boson interaction, making it suitable for studying the coupling of electrons with other quantized bosonic entities such as plasmons and phonons. As an example, our approach can be extended to the investigation of polaron formation from the first principles.

While VT-QEDHF is robust across all coupling strengths at the mean-field level, it inherently underestimates both intrinsic and photon-mediated electronic correlations. To address this limitation, our forthcoming research will focus on the integration of VT-QEDHF into QED-CCSD and EOM-CCSD frameworks. Given the superior performance of VT-QEDHF over existing QEDHF and SC-QEDHF methods, we are optimistic that the advanced QED-CC methods augmented with VT-QEDHF~\cite{haugland_coupled_2020, jcp0033132, Weight2023pccp, Liebenthal2022} will significantly improve correlation energy estimations in all coupling regimes.

{Additional note:} While drafting this manuscript, we became aware of a recent paper that employs similar concepts~\cite{Cui2023arxiv}. However, the variational transformation in Ref.~\cite{Cui2023arxiv} is limited to diagonal terms (of the dipole coupling operator) in the transformation. In contrast, our transformation is more general, and the corresponding elements are evaluated within the dipole basis.

\begin{acknowledgements}
We acknowledge support from the US DOE, Office of Science, Basic Energy Sciences, Chemical Sciences, Geosciences, and Biosciences Division under Triad National Security, LLC (``Triad") contract Grant 89233218CNA000001 (FWP: LANLECF7). This research used computational resources provided by the Institutional Computing (IC) Program and the Darwin testbed at Los Alamos National Laboratory (LANL), which is funded by the Computational Systems and Software Environments subprogram of LANL's Advanced Simulation and Computing program. LANL is operated by Triad National Security, LLC, for the National Nuclear Security Administration of the U.S. Department of Energy (Contract No. 89233218CNA000001). 
\end{acknowledgements}

{\bf Data availability.} The data supporting this study's findings are available from the corresponding author upon request.

{\bf Code availability.} The developed code used for this study is available from the corresponding author upon request. 

\bibliography{polaritonref}

\clearpage
\begin{widetext}    

\begin{center}
 {\Large\bf Supplementary Materials for ``First-principles molecular quantum electrodynamics theory at all coupling strengths"}
\end{center}

\setcounter{page}{1}
\setcounter{section}{0}
\setcounter{subsection}{0}
\setcounter{subsubsection}{0}
\setcounter{equation}{0}
\setcounter{figure}{0}
\renewcommand{\theequation}{S\arabic{equation}}
\renewcommand{\thesection}{S\arabic{section}}
\renewcommand\thefigure{S\arabic{figure}}

\section{QEDHF Method}
Like the Hartree-Fock (HF) method for purely electronic systems, the QEDHF reference wavefunction is a direct product of a single Slater determinant of electronic orbitals and a zero-photon state.

{\bf Fock State Representation.}
The QEDHF equations can be derived by starting with the reference states,
\begin{equation}\label{eq_pr}
    \ket{R}=\ket{\Psi_0}\otimes \left (\sum_{\boldsymbol{n}}C_{\boldsymbol{n}}\prod_{\alpha}\ket{n_\alpha} \right)
    \equiv \ket{\Psi_0}\otimes\ket{P}. 
\end{equation}
where $\boldsymbol{n}=(n_1, n_2, \ldots)$.
And $\ket{n_\alpha} = \frac{(\hat{a}^\dag_\alpha)^{n_\alpha}}{\sqrt{n_\alpha!}}\ket{0}$ are the normalized photon number states for mode $\alpha$. $\ket{\Psi_0}$ denotes the Slater determinant of electronic orbitals. For a given electronic state (such as an HF state), the total energy can then be minimized with respect to the photon coefficients $C_{\boldsymbol{n}}$, which is achieved by diagonalizing the Hamiltonian after integrating out the electronic degrees of freedom (DOF), resulting in the dressed photonic Hamiltonian,
\begin{align}\label{eq:photonicHam}
    \hH_P =& \bra{\Psi_0}\hH\ket{\Psi_0}
    \nonumber\\
    =& E_{M} + \sum_{\alpha} \Big[ \omega_\alpha (\ha^\dag_\alpha\ha_\alpha+\frac{1}{2}) + \frac{1}{2} \langle (\blambda_\alpha \cdot \hat{\boldsymbol{D}})^2 \rangle
    \nonumber\\
    &+\sqrt{\frac{\omega_\alpha}{2}} \blambda_\alpha \cdot \langle \hat{\boldsymbol{D}} \rangle (\ha^\dag_\alpha + \ha_\alpha) \Big].
\end{align}

In the evaluation of the expectation of the Dipole Self-Energy (DSE) operator, it should be noted that $\langle (\boldsymbol{\lambda}_\alpha \cdot \hat{\boldsymbol{D}})^2 \rangle \neq (\boldsymbol{\lambda}_\alpha \cdot \langle\hat{\boldsymbol{D}}\rangle)^2$ because
\begin{align}\label{eq:dseop1}
    & (\blambda_\alpha \cdot \hat{\boldsymbol{D}})^2
    =\blambda_\alpha \cdot \hat{\boldsymbol{D}} \blambda_\alpha \cdot \hat{\boldsymbol{D}}  
    =\sum_{\mu\nu\lambda\sigma}\bar{d}^\alpha_{\mu\nu}\bar{d}^\alpha_{\lambda\sigma}\hat c^\dag_\mu \hat c^\dag_\lambda \hat c_\sigma \hat c_\nu -\sum_{\mu\nu}\bar{q}^\alpha_{\mu\nu} \hat c^\dag_\mu \hat c_\nu.
\end{align}
where $\bar{d}^\alpha_{\mu\nu} = \boldsymbol{\lambda}_\alpha \cdot \bra{\mu}\hat{\boldsymbol{D}}\ket{\nu}$ and
$\bar{q}^\alpha_{\mu\nu} = \boldsymbol{\lambda}_\alpha \cdot \bra{\mu}\boldsymbol{q}\ket{\nu} \cdot \boldsymbol{\lambda}_\alpha$ are modified dipole and quadrupole integrals, respectively.
The above derivation does not assume the completeness of the one-particle basis set and is employed in Ref.~\cite{Vu2022, McTague2022}. Conversely, the second-quantized form for the square of the electric dipole operator is often approximated as the product of second-quantized electric dipole operators in many studies~\cite{foley:2023CPR},
\begin{align}
    (\blambda_\alpha \cdot \hat{\boldsymbol{D}})^2=&
    \sum_{\mu\nu}\bar{d}^\alpha_{\mu\nu}c^\dag_\mu c_\nu
    \sum_{\lambda\sigma}\bar{d}^\alpha_{\lambda\sigma}\hat c^\dag_\lambda \hat c_\sigma
    \nonumber\\
    =&\sum_{\mu\nu\lambda\sigma}\bar{d}^\alpha_{\mu\nu}\bar{d}^\alpha_{\lambda\sigma} \hat c^\dag_\mu \hat c^\dag_\lambda \hat c_\sigma \hat c_\nu 
    + \sum_{\mu\sigma} \left(\sum_\nu\bar{d}^\alpha_{\mu\nu}\bar{d}^\alpha_{\nu\sigma}\right) \hat c^\dag_\mu \hat c_\sigma.
\end{align}
Nevertheless, this expression shows that the DSE can be evaluated via photon-mediated one- and two-electron integrals. Consequently, the partial derivative of the QEDHF energy with respect to the electronic density matrix yields a new Fock matrix incorporating the DSE-mediated exchange and correlation matrix~\cite{foley:2023CPR}.
Alternatively, the total Fock matrix can be evaluated by modifying the one-electron and two-electron integrals,
\begin{align}
    h_{\mu\nu} & \rightarrow h_{\mu\nu} -\frac{1}{2}\sum_\alpha \tilde{q}^\alpha_{\mu\nu},
    \\
    I_{\mu\nu\lambda\sigma} &  \rightarrow I_{\mu\nu\lambda\sigma}  + \sum_\alpha \tilde{d}^\alpha_{\mu\nu} \tilde{d}^\alpha_{\lambda\sigma}
\end{align}

{\bf CS Representation}.
\label{sec:qedhf_cs}
In fact, Eq.~\ref{eq:photonicHam} can be diagonalized by the unitary transformation:
\begin{equation}\label{eq:u1}
    \hat{U}(\boldsymbol{z}) = \prod_\alpha \exp[z_\alpha \hat{a}^\dag_\alpha - z_\alpha^* \hat{a}_\alpha]
\end{equation}
where $z_\alpha = -\frac{\boldsymbol{\lambda}_\alpha \cdot \langle \hat{\boldsymbol{D}} \rangle}{\sqrt{2\omega_\alpha}}$.
The resulting PF Hamiltonian in the CS representation is
\begin{align}\label{eq:pfcs_ham}
    \hat{H}_{CS} = & H_e + \sum_\alpha \Big\{ \omega_\alpha \hat{a}^\dag_\alpha \hat{a}_\alpha
    +\frac{1}{2}[\boldsymbol{\lambda}_\alpha \cdot (\hat{\boldsymbol{D}} - \langle \hat{\boldsymbol{D}} \rangle)]^2
    \nonumber\\
    & -\sqrt{\frac{\omega_\alpha}{2}}[\boldsymbol{\lambda}_\alpha \cdot (\hat{\boldsymbol{D}} - \langle \hat{\boldsymbol{D}} \rangle)](\hat{a}^\dag_\alpha + \hat{a}_\alpha)\Big\}.
\end{align}
Note that
$\hat{U}^\dag \hat{a}_\alpha \hat{U} = \hat{a}_\alpha + [z^*_\alpha \hat{a}_\alpha - z_\alpha \hat{a}^\dag_\alpha, \hat{a}_\alpha] = \hat{a}_\alpha + z_\alpha.$

With the CS representation, the transformed Hamiltonian automatically ensures convergence with respect to the number of photon number states since a coherent state is a linear combination of many photon number states,
\begin{equation}
\ket{z_\alpha} \equiv \hat{U}(z_\alpha) \ket{0}
 = e^{-\frac{|z_\alpha|^2}{2}} \sum_{n=0}^{\infty} \frac{z_\alpha^n}{\sqrt{n!}} \ket{n_\alpha},
\end{equation}
where $\hat{U}(z_\alpha) = e^{-|z_\alpha|^2/2} e^{z_\alpha \hat{a}^\dag_\alpha} e^{-z^*_\alpha \hat{a}_\alpha}$ is used.

After the unitary transformation, the Hamiltonian can be solved with the ansatz
\begin{equation}
    \ket{R} = \ket{\text{HF}} \otimes \ket{0}.
\end{equation}
With this ansatz, the QEDHF energy is
\begin{align}\label{eq:eqedhf_cs}
    E_{\text{QEDHF}} = & E_{\text{HF}} + \frac{1}{2}\sum_\alpha \langle \boldsymbol{\lambda}_\alpha \cdot [\hat{\boldsymbol{D}} - \langle \hat{\boldsymbol{D}} \rangle)]^2 \rangle,
\end{align}
\textit{i.e.}, the bilinear coupling term in Eq.~\ref{eq:pfcs_ham} does not contribute to the QEDHF total energy when the Hamiltonian is represented in the coherent-state basis.
The electronic HF energy is $E_{\text{HF}} = \text{Tr}[h + \frac{1}{2}(J - K)]D$, where $D$ is the one-electron density matrix. $h$, $J$, and $K$ are the one-electron integral, Coulomb, and exchange potentials, respectively.
The DSE in CS representation becomes,
\begin{align}\label{eq:hdse_cs}
    \hat{H}_{\text{DSE}} \equiv & \frac{1}{2}\sum_\alpha \boldsymbol{\lambda}_\alpha \cdot [\hat{\boldsymbol{D}} - \langle \hat{\boldsymbol{D}} \rangle)]^2
    \nonumber\\
    = & \frac{1}{2}\sum_\alpha \left[(\boldsymbol{\lambda}_\alpha \cdot \hat{\boldsymbol{D}})^2 + (\boldsymbol{\lambda}_\alpha \cdot \langle \hat{\boldsymbol{D}} \rangle)^2
   - 2(\boldsymbol{\lambda}_\alpha \cdot \hat{\boldsymbol{D}})(\boldsymbol{\lambda}_\alpha \cdot \langle \hat{\boldsymbol{D}} \rangle)\right]
   \nonumber\\
   = & \frac{1}{2}\sum_\alpha \bar{d}^\alpha_{\mu\nu} \bar{d}^\alpha_{\lambda\sigma} \hat{c}^\dag_\mu \hat{c}^\dag_\lambda \hat{c}_\sigma \hat{c}_\nu
   - \sum_\alpha \left[\frac{1}{2} \bar{q}^\alpha_{\mu\nu} + (\boldsymbol{\lambda}_\alpha \cdot \langle \hat{\boldsymbol{D}} \rangle) \bar{d}^\alpha_{\mu\nu} \right] \hat{c}^\dag_\mu \hat{c}_\nu
   + \frac{1}{2}\sum_\alpha (\boldsymbol{\lambda}_\alpha \cdot \langle \hat{\boldsymbol{D}} \rangle)^2
\end{align}
where the first term is given by Eq.~\ref{eq:dseop1}. Hence, substituting Eq.~\ref{eq:hdse_cs} into the QEDHF energy expression in CS representation (Eq.~\ref{eq:eqedhf_cs}) is equivalent to an electronic HF energy with modified two-electron and one-electron integrals, subject to a difference of $\frac{1}{2}\sum_\alpha (\boldsymbol{\lambda}_\alpha \cdot \langle \hat{\boldsymbol{D}} \rangle)^2$,
\begin{align}
  h_{\mu\nu} \rightarrow & h_{\mu\nu} - \sum_\alpha \left[\frac{1}{2}\bar{q}^\alpha_{\mu\nu} + (\boldsymbol{\lambda}_\alpha \cdot \langle \hat{\boldsymbol{D}} \rangle) \bar{d}^\alpha_{\mu\nu} \right],
  \\
  I_{\mu\nu\lambda\sigma} \rightarrow & I_{\mu\nu\lambda\sigma} + \sum_\alpha \bar{d}^\alpha_{\mu\nu} \bar{d}^\alpha_{\lambda\sigma}.
\end{align}

\section{Variational QED-HF Theory for Arbitrary Coupling Strength}

In this section, we describe the variational polaron transformation-based QED-HF method.
The parameterized unitary transformation is defined as
\begin{align}
    \hat{U}_f
    &= e^{-\sum_\alpha \frac{f_\alpha}{\sqrt{2\omega_\alpha}} \boldsymbol{e}_\alpha\cdot\hat{\boldsymbol{D}}(\hat{a}_\alpha-\hat{a}^\dag_\alpha)}.
\end{align}
where $\boldsymbol{f}=\{f_\alpha\}$ are the parameters to be optimized.
The electronic and photonic operators, after the transformation, become
\begin{align}
    \hat{U}^\dag_f \hat{c}_\nu \hat{U}_f &= \sum_\mu c_\nu X_{\mu\nu}, \\
    \hat{U}^\dag_f \hat{a}_\alpha \hat{U}_f &= \hat{a}_\alpha - \frac{f_\alpha}{\sqrt{2\omega_\alpha}} \boldsymbol{e}_\alpha\cdot\hat{\boldsymbol{D}}.
\end{align}
where
\begin{equation}
    X_{\mu\nu} =\exp\left[-\sum_{\alpha}\frac{f_\alpha}{\sqrt{2\omega_\alpha}} \boldsymbol{e}_\alpha\cdot\hat{\boldsymbol{D}}(\hat{a}^\dag_\alpha - \hat{a}_\alpha) \right]|_{\mu\nu}.
\end{equation}
Note that the electronic operator is not present in $X_{\mu\nu}$ in contrast to the unitary operator $\hat{U}_f$. To derive the above equations, we used the identity $e^S A e^{-S}=\sum_n \frac{1}{n!}A^{(n)}$ where $A^{(n)}=[S,A^{(n-1)}]$ and $A^{(0)}=A$. For the electronic operator $\hat{c}_\mu$, we have (we define $\zeta^\alpha_{\mu\nu}=\frac{f_\alpha}{\sqrt{2\omega_\alpha}} (\boldsymbol{e}_\alpha\cdot\hat{\boldsymbol{D}})_{\mu\nu}(\hat{a}_\alpha-\hat{a}^\dag_\alpha)$),
\begin{align}
    \hat{c}^{(1)}_\mu =& [S, \hat{c}_\mu]=\sum_{\alpha,\mu'\nu}\zeta^\alpha_{\mu'\nu}[\hat{E}_{\mu'\nu}, \hat{c}_\mu] = -\sum_{\alpha,\nu}\zeta^\alpha_{\mu\nu} \hat{c}_\nu, \\
    \hat{c}^{(2)}_\mu = & [S, \hat{c}^{(1)}_\mu]=\sum_{\alpha,\mu'\nu'}\zeta^\alpha_{\mu'\nu'}[\hat{E}_{\mu'\nu'}, -\sum_{\nu}\zeta^{\alpha}_{\mu\nu}\hat{c}_\nu]
    = \sum_{\alpha\nu\nu'}\zeta^\alpha_{\mu\nu}\zeta^\alpha_{\nu\nu'}\hat{c}_{\nu'}=-\sum_\nu(\zeta^\alpha)^2_{\mu\nu} \hat{c}_\nu
\end{align}
and so on. After the transformation, the Hamiltonian becomes
\begin{align}
      \hat{H}(\{f_\alpha\}) = & \tilde{H}_e +
      \sum_{\alpha} \omega_{\alpha}\left(\hat{a}^\dag_\alpha - \frac{f_\alpha}{\sqrt{2\omega_\alpha}}  \boldsymbol{e}_\alpha\cdot\hat{\boldsymbol{D}}\right)\left(\hat{a}_\alpha - \frac{f_\alpha}{\sqrt{2\omega_\alpha}} \boldsymbol{e}_\alpha\cdot\hat{\boldsymbol{D}}\right) \nonumber\\
      & + \sum_{\alpha}\sqrt{\frac{\omega_\alpha}{2}}\lambda_\alpha  \boldsymbol{e}_\alpha\cdot\hat{\boldsymbol{D}}\left(\hat{a}^\dag_\alpha+\hat{a}_\alpha  - 2 \frac{f_\alpha}{\sqrt{2\omega_\alpha}} \boldsymbol{e}_\alpha\cdot\hat{\boldsymbol{D}}\right) + \frac{\lambda^2_\alpha}{2}(\boldsymbol{e}_\alpha\cdot\hat{\boldsymbol{D}})^2 \nonumber\\
      =&\tilde{H}_e + \sum_{\alpha} \omega_{\alpha}\hat{a}^\dag_\alpha \hat{a}_\alpha
      + \sum_{\alpha}\sqrt{\frac{\omega_\alpha}{2}}(\Delta\lambda_\alpha) \boldsymbol{e}_\alpha\cdot\hat{\boldsymbol{D}}(\hat{a}^\dag_\alpha+\hat{a}_\alpha )
      +  \frac{(\Delta\lambda_\alpha)^2}{2}(\boldsymbol{e}_\alpha\cdot\hat{\boldsymbol{D}})^2.
\end{align}
which is the Eq.~\ref{eq:varh} in the main text and $\Delta\lambda_\alpha=\lambda_\alpha-f_\alpha$. Hence, after the variational transformation, Eq.~\ref{eq:varh} is formally the same as the original Hamiltonian with $1)$ $\lambda_\alpha$ replaced with $\Delta\lambda_\alpha$ and $2)$ photonic displacement operator dresses electronic integrals. The dressed electronic Hamiltonian reads
\begin{align}
    \tilde{H}_e = \tilde{h}_{\mu\nu}\hat{c}^\dag_\mu \hat{c}_\nu + \tilde{I}_{\mu\nu\lambda\sigma}\hat{c}^\dag_\mu \hat{c}^\dag_\lambda \hat{c}_\sigma \hat{c}_\nu.
\end{align}
where
\begin{align}
    \tilde{h}_{\mu\nu}&=\sum_{\mu'\nu'} h_{\mu'\nu'}X^\dag_{\mu\mu'}X_{\nu\nu'}, \\
    \tilde{I}_{\mu\nu\lambda\sigma}&= \sum_{\mu'\nu'\lambda'\sigma'}X^\dag_{\mu\mu'}X^\dag_{\lambda\lambda'}I_{\mu'\nu'\lambda'\sigma'} X_{\nu\nu'}X_{\sigma\sigma'}
\end{align}
are the dressed one- and two-electron integrals.

\subsubsection{QED energies and dipole basis}                               
The evaluation of displacement operator elements depends on the diagonalization of the $\boldsymbol{e}_\alpha\cdot\hat{\boldsymbol{D}}$ matrix. Therefore, transforming the basis into a dipole basis set simplifies the process,    
\begin{align}\label{eq:dipole}
    X_{\mu\nu} =\prod_\alpha & V^\alpha_{\mu p} \exp\left[-\frac{f_\alpha}{\sqrt{2\omega_\alpha}} (\boldsymbol{e}_\alpha\cdot\hat{\boldsymbol{D}})_{p}(\ha^\dag_\alpha - \ha_\alpha) \right]V^\alpha_{p\nu}.
\end{align}                                                                 
where $V^\alpha$ is the transformation matrix that diagonalizes the dipole coupling matrix $(\boldsymbol{e}_\alpha\cdot\hat{\boldsymbol{D}})_{\mu\nu}$. Thus, we can rewrite the original Hamiltonian in the dipole basis as introduced in Ref.~\onlinecite{Riso:2022uw}.

Consequently, the one-electron part of the QEDHF energy is $E_T =\sum_{\mu\nu}\tilde{h}_{\mu\nu}\rho_{\mu\nu}$, where the photon-dressed one-electron integral is
\begin{align}                                                               
    \bra{0_p}\tilde{h}_{\mu\nu}\ket{0_p}=& \left[\sum_{\mu'\nu'} \tilde{h}_{\mu'\nu'} X^\dag_{\mu\mu'}X_{\nu\nu'}\right]
    \nonumber\\                                                             
    =&\left[\sum_{\mu'\nu'} \sum_{pq}h_{\mu'\nu'} \prod_\alpha U^\alpha_{\mu p} e^{\frac{f_\alpha}{\sqrt{2\omega_\alpha}} (\boldsymbol{e}_\alpha\cdot\hat{\boldsymbol{D}})_{p}(\ha^\dag_\alpha - \ha_\alpha)} U^\alpha_{p\mu'}  U^\alpha_{\nu q} e^{-\frac{f_\beta}{\sqrt{2\omega_\beta}} \boldsymbol{e}_\alpha\cdot\hat{\boldsymbol{D}}_{q}(\ha^\dag_\beta - \ha_\beta)}U^\alpha_{q\nu'}\right]
    \nonumber\\
    =&\left[\sum_{\mu'\nu'} \sum_{pq}\tilde{h}_{\mu'\nu'} \prod_\alpha U^\alpha_{\mu p} U^\alpha_{\nu q} 
    G^\alpha_{pq} U^\alpha_{p\mu'} U^\alpha_{q\nu'}\right].
\end{align}
The two-electron integrals in the photonic vacuum state are
\begin{align}
    \bra{0_p}\tilde{I}_{\mu\nu\lambda\sigma}\ket{0_p}=& \sum_{\mu'\nu'\lambda'\sigma'}X^\dag_{\mu\mu'}X^\dag_{\nu\nu'}I_{\mu'\nu'\lambda'\sigma'} X_{\lambda\lambda'}X_{\sigma\sigma'}
    \nonumber\\
    =&\sum_{\mu'\nu'\lambda'\sigma'}\sum_{pqrs} \prod_{\alpha} V^\alpha_{\mu'p}V^\alpha_{p\mu}V^\alpha_{\nu' q}V_{q\nu} 
    I_{\mu'\nu'\lambda'\sigma'}
    V^\alpha_{\lambda' r} V^\alpha_{r\lambda}V^\alpha_{\sigma's}V^\alpha_{s\sigma}G^\alpha_{pqrs}.
\end{align}
Where the Gaussian factors are
\begin{align}
    G^\alpha_{pq}= & \exp\left[-\sum_\alpha\frac{f^2_\alpha(\eta^\alpha_p - \eta^\alpha_q)^2}{4\omega_\alpha} \right],
    \\
    G^\alpha_{pqrs}= & \exp\left[-\sum_\alpha\frac{f^2_\alpha(\eta^\alpha_p - \eta^\alpha_q+\eta^\alpha_r - \eta^\alpha_s)^2}{4\omega_\alpha} \right].
\end{align}

Additionally, the DSE residue contributes to the total energy as well. Analogous to the QED-HF formalism, the residual DSE contribution can be computed via DSE-mediated one-electron and two-electron integrals 
\begin{align}
   h^p_{\mu\nu}=& -\sum_\alpha\frac{(\Delta\lambda_\alpha)^2}{2}\tilde{q}_{\mu\nu},
   \\
   I^p_{\mu\nu\lambda\sigma}=&\sum_\alpha\frac{(\Delta\lambda_\alpha)^2}{2} \tilde{d}_{\mu\nu}\tilde{d}_{\lambda\sigma}.
\end{align}
This formulation allows the residual DSE-mediated Fock matrix and its associated energies to be expressed in a manner analogous to the electronic components.

In summary, the total energy in the dipole basis is given by
\begin{align}
    E = &
    \sum_{pq} \tilde{h}_{pq}\rho_{pq}G_{pq}
    + \frac{1}{2}\sum_{pqrs}\tilde{I}_{pqrs}\left(\rho_{pq}\rho_{rs} - \frac{1}{2}\rho_{ps}\rho_{rq}\right) G_{pqrs}
    + \sum_{\alpha p} \frac{f^2_\alpha}{2} \rho_{pp}[(\boldsymbol{e}_\alpha\cdot\hat{\boldsymbol{D}})_{pp} - \eta_p]^2
    \nonumber\\
    & + \sum_{pq\alpha}\frac{f^2_\alpha}{2} \left(\rho_{pp}\rho_{qq}-\frac{1}{2}\rho_{pq}\rho_{qp}\right)[(\boldsymbol{e}_\alpha\cdot\hat{\boldsymbol{D}})_{pp} - \eta_p][(\boldsymbol{e}_\alpha\cdot\hat{\boldsymbol{D}})_{qq} - \eta_q]
    \nonumber\\
    & + \sum_\alpha\frac{(\Delta\lambda_\alpha)^2}{2}\bra{\HF}\bra{0}(\boldsymbol{e}_\alpha\cdot\hat{\boldsymbol{D}})^2\ket{\HF}\ket{0}.
\end{align}

\subsubsection{Gradients of total energy with respect to variational transformation parameter}
The variational optimization of the $\{f_\alpha\}$ parameters is achieved via the variational minimization procedure along with the density matrix optimization. In particular, the optimal transformation parameters are obtained when the energy gradients with respect to the $\{f_\alpha\}$ are equal to zero. The gradient (in the dipole basis) is given by
\begin{align}
    \frac{\partial E}{\partial f_\alpha}= & \frac{\partial E_e}{\partial f_\alpha} + \frac{\partial E_{DSE}}{\partial f_\alpha}
    \nonumber\\
    =& \sum_{pq} \frac{-f_\alpha (\eta_{\alpha,p}-\eta_{\alpha,q})^2}{2\omega_\alpha} \tilde{h}_{pq}\rho_{pq} G_{pq} 
    \nonumber\\
    & + \sum_{pqrs} \frac{-f_\alpha(\eta^\alpha_p - \eta^\alpha_q + \eta^\alpha_r - \eta^\alpha_s)^2}{2\omega_\alpha} \tilde{I}_{pqrs}\left(\rho_{pq}\rho_{rs} - \frac{1}{2}\rho_{ps}\rho_{rq}\right) G_{pqrs}
    \nonumber\\
    & + f_\alpha \sum_p \rho_{pp}[(\boldsymbol{e}_\alpha\cdot\hat{\boldsymbol{D}})_{pp} - \eta_p]^2
    \nonumber\\
    & + f_\alpha \sum_{pq}\left(\rho_{pp}\rho_{qq}-\frac{1}{2}\rho_{pq}\rho_{qp}\right)[(\boldsymbol{e}_\alpha\cdot\hat{\boldsymbol{D}})_{pp} - \eta_p][(\boldsymbol{e}_\alpha\cdot\hat{\boldsymbol{D}})_{qq} - \eta_q]
    \nonumber\\
    & -  \Delta\lambda_\alpha \bra{\HF}\bra{0}(\boldsymbol{e}_\alpha\cdot\hat{\boldsymbol{D}})^2\ket{\HF}\ket{0}.
\end{align}

\end{widetext}

\end{document}